\documentclass[nojss]{jss}\usepackage[]{graphicx}\usepackage[]{color}
\makeatletter
\def\maxwidth{ %
  \ifdim\Gin@nat@width>\linewidth
    \linewidth
  \else
    \Gin@nat@width
  \fi
}
\makeatother

\usepackage{graphicx,epsfig,amsmath,alltt,setspace,bm,float,tabularx}
\PassOptionsToPackage{hyphens}{url}
\usepackage[english]{babel}
\usepackage{tikz}
\usetikzlibrary{arrows}
\usepackage[parfill]{parskip}

\title{\pkg{merDeriv}: Derivative Computations for Linear Mixed 
  Effects Models with Application to Robust Standard Errors}

\author{Ting Wang\\The American Board of Anesthesiology
  \And Edgar C.\ Merkle \\University of Missouri}
\Plainauthor{Ting Wang, Edgar C.\ Merkle}

\Shorttitle{Derivative Computations for Linear Mixed Effects Models}
\Plaintitle{Accepted at Journal of Statistical Software}

\Address{
  Ting Wang\\
  The American Board of Anesthesiology\\
  4208 Six Forks Road, Suite 1500\\
  Raleigh, NC 27609-5765\\
  E-mail: \email{ting.wang@theaba.org}
  \\
  \\
  Edgar C.\ Merkle\\
  Department of Psychological Sciences\\
  University of Missouri\\
  28A McAlester Hall\\
  Columbia, MO, USA 65211\\
  E-mail: \email{merklee@missouri.edu}\\
  URL: \url{http://faculty.missouri.edu/~merklee/}
}

\Abstract{While likelihood-based derivatives and related facilities 
  are available
  in \proglang{R} for many types of statistical models, the facilities are
  notably lacking for models estimated via \pkg{lme4}. This is because the
  necessary statistical output, including the Hessian, Fisher information 
  and casewise contributions to the model gradient, is not immediately 
  available from \pkg{lme4} and is
  not trivial to obtain. In this article, we describe \pkg{merDeriv}, an 
  \proglang{R} package which supplies new 
  functions to obtain \emph{analytic} output from Gaussian mixed models. 
  We discuss the theoretical results implemented in the code, 
  focusing on calculation of robust standard errors via package 
  \pkg{sandwich}. We also use the \code{sleepstudy} data to illustrate the 
  package and to compare it to a benchmark from package \pkg{lavaan}.}  
\Keywords{Linear mixed effects model, scores, Huber-White sandwich estimator,
  robust standard error, \pkg{lme4}}

\spacing{1}
\IfFileExists{upquote.sty}{\usepackage{upquote}}{}
\begin{document}

\section{Introduction}
Package \pkg{lme4} \citep{lme4} is widely used to estimate a variety of
generalized linear mixed models.  Despite its popularity, the package does
not provide certain results related to derivatives of the likelihood,
which makes it difficult to obtain robust standard errors and other
statistical tests.  This absence is partially related to the
fact that \pkg{lme4} does not directly estimate models via likelihood
maximization, but rather employs a penalized least squares approach
that leads to ML (or REML) estimates \citep{lme4}.  While this approach
eases model estimation, it also makes it more difficult to obtain
derivatives (first and second) of the likelihood from a
fitted model (which are required for, e.g., the Huber-White sandwich 
estimator).  While it is possible to instead utilize
the robust estimation methods from package \pkg{robustlmm} \citep{roblmm},
we are interested in directly using derivative-based methods that rely on 
estimation of the traditional model. Thus, the goal of this paper is to 
describe 
\proglang{R} package \pkg{merDeriv}, which contains functions that 
compute these derivatives for objects of class \code{lmerMod}.  
We also briefly discuss derivatives associated with models of 
class \code{glmerMod}, though we do not currently have code for these 
models (the computations are more difficult due to 
the need for numerical integration).

The paper proceeds as follows.
We first describe general notation for the linear mixed model.  Next, we derive
expressions for the linear mixed models' casewise (observation level) and 
clusterwise (cluster/group level) first 
derivatives, along with the Hessian and Fisher information 
matrix (including both fixed effect parameters and variances/covariances 
of random effects. 
Next, we illustrate the derivatives' application via 
the sleep study data \citep{belenky03} included with \pkg{lme4}, comparing 
our results to a benchmark from \pkg{lavaan} \citep{rosv12}.
This illustration includes computation of the Huber-White sandwich 
estimator \citep{eicker67, white80, huber67} for 
linear mixed models with independent clusters/groups.
Finally, we discuss further use and extension of our package's functionality.

\section{Linear mixed model}
%Let $\bm y$ include $n$ observations clustered in $J$ groups.
%$c_j=1, 2, \ldots, J$ groups.
Following \cite{lme4}, the linear mixed model can be written as
\begin{eqnarray}
  \label{eq:lmmcond}
    \bm y |\bm b &\sim& N(\bm X \bm \beta+\bm Z\bm b, \bm R)\\
  \label{eq:lmmran}
    \bm b &\sim& N(\bm 0, \bm G)\\
  \label{eq:lmmres}
    \bm R &=& \sigma_{r}^2\bm I_{n},
\end{eqnarray}
where $\bm y$ is the observed data vector of length $n$; $\bm X$ is an
$n \times p$ matrix of fixed covariates; $\bm \beta$ is the fixed effect
vector of length $p$; $\bm Z$ is an $n \times q$ design matrix of random
effects; and $\bm b$ is the random effect vector of
length $q$.

The vector $\bm b$ is assumed to follow a normal distribution with
mean $\bm 0$ and covariance matrix $\bm G$, where $\bm G$ is a block diagonal
matrix composed of varaiance/covariance for random effect parameters.
The residual covariance matrix, $\bm R$, is the product of the residual
variance $\sigma_{r}^2$ and an identity matrix of dimension $n$.
We further define $\bm \sigma^2$ to be a vector of length $K$, containing all
variance/covariance parameters (including those of the random
effects and the residual).  Thus, the matrix $\bm G$ has $(K-1)$ 
unique elements.  For example, in a model with two random effects 
that are allowed to covary, $\bm \sigma^{2}$ is a vector of 
length 4 (i.e., $K = 4$).  The first three elements correspond to the unique
entries of $\bm G$, which are commonly expressed as 
$\sigma_0^2$, $\sigma_0\sigma_1$, and $\sigma_1^2$.  
The last component is then the residual variance $\sigma_r^2$.

Based on Equations~\ref{eq:lmmcond}, \ref{eq:lmmran},
and \ref{eq:lmmres}, the marginal distribution of the LMM is
\begin{equation}
  \label{eq:marginml}
  \bm y \sim N(\bm X \bm \beta, \bm V),
\end{equation}
where
\begin{equation}
  \label{eq:marginv}
    \bm V = \bm Z \bm G \bm Z^{\top} + \sigma_{r}^2\bm I_{n}.
\end{equation}
Therefore, the marginal likelihood can be expressed as
\begin{equation}
  \label{eq:obj}
    \ell(\bm \sigma^2, \bm \beta; \bm y) = -\frac{n}{2}\log(2\pi) -
    \frac{1}{2}\log(|\bm V|) - \frac{1}{2}
    (\bm y - \bm X \bm \beta)^{\top}\bm V^{-1} (\bm y- \bm X \bm \beta).
\end{equation}

\section{Derivative computations for the linear mixed model}
In this section, we first discuss analytic results involving the linear
mixed model's first and second derivatives.  We then illustrate how
these derivatives can be obtained from an object of class \code{lmerMod}.

\subsection{Scores}
Based on the objective function from Equation~\ref{eq:obj},
we derive the score function $s_i()$ for each observation w.r.t.\ the
parameter vector $\bm \xi = (\bm \sigma^2, \bm \beta)^\top$.  We focus
separately on $\bm \sigma^2$ and on $\bm \beta$ below.

\subsubsection[Random scores]{Scores for $\bm \sigma^2$}
The gradient with respect to the $k^{\text{th}}$ entry
of $\bm \sigma^2$ ($k=1, 2, 3, \ldots, K$) is \cite[p.\ 136--137]{stroup12}:
\begin{equation}
  \label{eq:grasigma}
    \frac{\partial \ell(\bm \sigma^2, \bm \beta; \bm y)}
         {\partial \sigma_k^2} = -\frac{1}{2}
      \text{tr} \left [\bm V^{-1} \frac{\partial \bm V}
        {\partial \sigma_k^2}\right ] +
        \frac{1}{2}(\bm y-\bm X \bm \beta)^{\top} \bm V^{-1} 
        \left (\frac{\partial \bm V}{\partial \sigma_k^2}\right ) 
        \bm V^{-1}(\bm y- \bm X \bm \beta),
\end{equation}
where $\bm V$ is defined in Equation~\ref{eq:marginv}. This gradient sums over 
$i$, whereas the scores are defined for each observation $i$.
Thus, to obtain the scores, we can remove the sums from the above equation. 
This is accomplished by replacing a trace operator with a diag operator, 
as well as replacing a matrix product with a Hadamard product (also known as 
elementwise/entrywise multiplication):
\begin{equation}
  \label{eq:scoresigma}
    s(\sigma_k^2; \bm y)= -\frac{1}{2}
    \text{diag} \left [\bm V^{-1} \frac{\partial \bm V}
      {\partial \sigma_k^2}\right ] +
    \left \{\frac{1}{2}(\bm y-\bm X \bm \beta)^{\top} \bm V^{-1} \left
    (\frac{\partial \bm V}{\partial \sigma_k^2}\right ) \bm V^{-1}\right\}^{T}
    \circ (\bm y- \bm X \bm \beta).
\end{equation}
In this way, the gradient  
of parameter $\sigma_k^2$ (a scalar) becomes a $n \times 1$ score vector.

\subsubsection[Fixed scores]{Scores for $\bm \beta$}
For the fixed effect parameter $\bm \beta$, the gradient is:
\begin{equation}
  \label{eq:grabeta}
  \frac{\partial \ell(\bm \sigma^2, \bm \beta; \bm y)}{\partial \bm \beta}
   = \bm X^{\top} \bm V^{-1}(\bm y-\bm X \bm \beta).
\end{equation}
The score vector $s(\bm \beta; \bm y)$ can again be obtained by
replacing the matrix multiplication by the Hadamard product:
\begin{equation}
  \label{eq:scoresbeta}
    s(\bm \beta; \bm y)
    = \left\{\bm X^{\top} \bm V^{-1} \right\}^{T} \circ (\bm y-\bm X
      \bm \beta).
\end{equation}
The full set of scores can then be
expressed as a matrix whose columns consist of the results from
Equations~\ref{eq:scoresigma} and \ref{eq:scoresbeta}.

These equations provide scores for each observation $i$, and we can
construct the clusterwise scores by summing scores within each cluster.
In situations with one grouping (clustering) variable, the clusterwise 
scores can be obtained from our \code{estfun.lmerMod()} function via the 
default argument \code{level = 2}.  % In situations with
% multiple grouping variables (i.e., crossed random effects, 
% three-level models), the function returns an error (more detail in the 
% Discussion section).
The casewise scores, on the other hand, can be
retrieved for all models via the argument \code{level = 1}.

\subsection{Hessian/observed information matrix}
The Hessian is the second derivative of the log-likelihood, 
noted as $A^{\star}$ in this paper.  The negative of the
Hessian is often called the observed information matrix or observed 
Fisher information.  It is a sample-based version of the Fisher information. Because
package \pkg{lme4} does not
provide a Hessian that includes both the fixed and 
variance/covariance of 
random effect (including residual variance)
parameters, the derivation of this matrix requires special attention.

To obtain the Hessian, we can divide the matrix $\bm A^{\star}$ into the 
following four blocks:

$$
        \bm A^{\star} = \left[\begin{array}{ccc|ccc}
                   &&\\
                   &\frac{\partial^2 \ell(\bm \sigma^2, \bm \beta; 
                      \bm y)}
                   {\partial \bm \beta \partial \bm \beta^{T}} &&&
                     \frac{\partial^2 \ell(\bm \sigma^2, \bm \beta; 
                       \bm y)}
                   {\partial \bm \beta \partial \bm \sigma^2}&\\
                   &&\\
                   \hline
                   &&\\
                   &\frac{\partial^2 \ell(\bm \sigma^2, \bm \beta; 
                     \bm y)}
                   {\partial \bm \sigma^2 \partial \bm \beta} &&&
                    \frac{\partial^2 \ell(\bm \sigma^2, \bm \beta; 
                      \bm y)}
                   {\partial \bm \sigma^2 \partial \bm \sigma^2} &\\
                   &&\\
                   \end{array}\right],
$$

where $\bm \beta$ contains all fixed parameters and $\bm \sigma^2$
contains all variance-covariance parameters
(in variance-covariance scale) in the linear mixed model.  To facilitate 
the analytic derivations, we index the above four blocks as:

$$
        \bm A^{\star} = \left[\begin{array}{ccc|ccc}
                   &&\\
                   &\text{Block 1}^{\star} &&& \text{Block 3}^{\star} &\\
                   &&\\
                   \hline
                   &&\\
                   &\text{Block 2}^{\star} &&& \text{Block 4}^{\star}&\\
                   &&\\
                   \end{array}\right].
$$

$\text{Block 1}^{\star}$ is straightforward, which can be obtained by taking 
the derivative of 
Equation~\ref{eq:grabeta} w.r.t.\ $\beta$, which can be expressed as: 
\begin{equation}
  \label{eq:hessianbeta}
    \frac{\partial^2 \ell(\bm \sigma^2, \bm \beta; \bm y)}
    {\partial \bm \beta \partial \bm \beta^{T}}  = -
    \bm X^{\top} \bm V^{-1} \bm X
\end{equation}

Derivation of $\text{Block 4}^{\star}$ is described in \cite{stroup12} and 
can be written as
\begin{multline}
  \label{eq: hessianrandom}
    \frac{\partial^2 \ell(\bm \sigma, \bm y, \bm \beta)}
      {\partial \sigma_{k_1}^2\sigma_{k_2}^2} =
      \left (\frac{1}{2}\right) \text{tr}
      \left [
      \bm V^{-1} \left (\frac{\partial \bm V}
      {\partial \sigma_{k1}}\right) \bm V^{-1} \left (\frac{\partial \bm V}
      {\partial \sigma_{k2}}\right) \right ]\\ 
      - (\bm y- \bm X \bm \beta)\left\{\bm V^{-1} \left (\frac{\partial \bm V}
      {\partial \sigma_{k1}}\right) \bm V^{-1} \left (\frac{\partial \bm V}
      {\partial \sigma_{k2}}\right) \bm V^{-1}\right\}\left(\bm y- \bm X 
      \bm \beta \right),
\end{multline}
where $k_1 \in 1, \ldots, K$ and $k_2 \in 1, \ldots, K$.

Finally, $\text{Block 3}^{\star}$ (which is the transpose of 
$\text{Block 2}^{\star}$) can
be seen as the derivative of Equation~\ref{eq:grasigma} 
w.r.t.\ $\bm \beta$. 
Using the identity from \citet[p.\ 11, Eq.\ (86)]{peter08}, 
this allows us to derive $\text{Block 3}^{\star}$ as
\begin{equation}
  \label{eq:covbeta}
    \frac{\partial^2 \ell(\bm \sigma^2, \bm \beta; \bm y)}
    {\partial \bm \sigma^2 \partial \bm \beta} = -\bm
    X^{\top} \bm V^{-1} \left ( \frac{\partial \bm V}
    {\partial \bm \sigma^2} \right )
    \bm V^{-1}(\bm y- \bm X \bm \beta) 
\end{equation}
The results obtained for $\bm A^{\star}$ are similar to the derivation 
of Fisher information matrix, as described below.

\subsection{Fisher/expected information matrix}
The Fisher information matrix (or expected information matrix) 
 is the expectation of the negative second 
derivative of the 
log likelihood, noted as 
$\bm A$ throughout the paper. It 
%$\bm A$ matrix in Equation~\ref{eq:aba},
can often be obtained in \proglang{R} with the help of the
\code{vcov()} function, but package \pkg{lme4} only 
provides results for fixed effect parameters.  Thus, we obtain the 
Fisher information 
w.r.t. all model 
parameters by taking the expectation of the negative of the Hessian 
matrix $\bm A^{\star}$.

Specifically, we can express the matrix $\bm A$ in the 
following four blocks as before.  The only difference is the negative 
expectation operator. 

$$
        \bm A = \left[\begin{array}{ccc|ccc}
                   &&\\
                   &-E \left (\frac{\partial^2 \ell(\bm \sigma^2, \bm \beta; 
                      \bm y)}
                   {\partial \bm \beta \partial \bm \beta^{T}} \right ) &&&
                     -E \left (\frac{\partial^2 \ell(\bm \sigma^2, \bm \beta; 
                       \bm y)}
                   {\partial \bm \beta \partial \bm \sigma^2} \right ) &\\
                   &&\\
                   \hline
                   &&\\
                   &-E \left (\frac{\partial^2 \ell(\bm \sigma^2, \bm \beta; 
                     \bm y)}
                   {\partial \bm \sigma^2 \partial \bm \beta} \right ) &&&
                    -E \left (\frac{\partial^2 \ell(\bm \sigma^2, \bm \beta; 
                      \bm y)}
                   {\partial \bm \sigma^2 \partial \bm \sigma^2} \right )&\\
                   &&\\
                   \end{array}\right],
$$

Following the same strategy, we index the above four blocks as:

$$
        \bm A = \left[\begin{array}{ccc|ccc}
                   &&\\
                   &\text{Block 1} &&& \text{Block 3}&\\
                   &&\\
                   \hline
                   &&\\
                   &\text{Block 2} &&& \text{Block 4}&\\
                   &&\\
                   \end{array}\right].
$$

Because $\bm X$ and $\bm X^{T}$ are considered  
constants, Block 1 is simply the negative of $\text{Block 1}^{\star} $ shown as 
below: 
\begin{equation}
  \label{eq:infobeta}
    -E \left (\frac{\partial^2 \ell(\bm \sigma^2, \bm \beta; \bm y)}
    {\partial \bm \beta \partial \bm \beta^{T}} \right ) = -E \left (-
    \bm X^{\top} \bm V^{-1} \bm X \right ) =  \bm X^{\top} \bm V^{-1} \bm X. 
\end{equation}
This analytic result is mathematically equivalent to the result 
provided by \code{solve(vcov())} in \pkg{lme4} (which only contains fixed
effect parameters).

Derivation of Block 4 is also based on the result 
from $\text{Block 4}^{\star}$.  In
particular, following the expectation identity for Gaussian distributions from 
\citet[p.\ 43, Eq.\ (380)]{peter08},
the second term of Equation~\ref{eq: hessianrandom} can be 
transformed to $\text{tr}\left[ \bm V^{-1} \left (\frac{\partial \bm V}
{\partial \sigma_{k1}^2}\right)
\bm V^{-1} \left (\frac{\partial \bm V}
{\partial \sigma_{k2}^2}\right) \right]$.  Thus Block 4 is reduced to 
the form shown as below, which is also described in \cite{stroup12}. 
\begin{equation}
  \label{eq: inforandom}
    -E \left (\frac{\partial^2 \ell(\bm \sigma, \bm y, \bm \beta)}
      {\partial \sigma_{k_1}^2\sigma_{k_2}^2} \right ) =
      \left (\frac{1}{2}\right) \text{tr}
      \left [ \bm V^{-1} \left (\frac{\partial \bm V}
      {\partial \sigma_{k1}^2}\right)
      \bm V^{-1} \left (\frac{\partial \bm V}
      {\partial \sigma_{k2}^2}\right)\right ],
\end{equation}
where $k_1 \in 1, \ldots, K$ and $k_2 \in 1, \ldots, K$.

Finally, Block 3 is the negative of the expectation of $\text{Block 3}^{\star}$. 
Using the expectation identity from \citet[p.\ 35, Eq.\ (312)]{peter08},
this allows us to derive Block 3 as
\begin{eqnarray}
  \label{eq:infocovbeta}
    -E \left (\frac{\partial^2 \ell(\bm \sigma^2, \bm \beta; \bm y)}
    {\partial \bm \sigma^2 \partial \bm \beta}\right ) &=& -E \left (-\bm
    X^{\top} \bm V^{-1} \left ( \frac{\partial \bm V}
    {\partial \bm \sigma^2} \right )
    \bm V^{-1}(\bm y- \bm X \bm \beta) \right) \\
    & = & \bm
    X^{\top} E \left (\left\{ \bm V^{-1} \left ( \frac{\partial \bm V}
    {\partial \bm \sigma^2} \right )
    \bm V^{-1} \right \} \left\{(\bm y- \bm X \bm \beta) \right\}\right)
\end{eqnarray}
Since $E(\bm y- \bm X \bm \beta) = \bm 0$,  it leads to 
$-E \left (\frac{\partial^2 \ell(\bm \sigma^2, \bm \beta; \bm y)}
{\partial \bm \sigma^2 \partial \bm \beta}\right ) = \bm 0$, which 
reflects
asymptotic independence of $\bm \beta$ and $\bm \sigma^2$. 
Thus, we have expressed the necessary derivatives as functions of model
matrices and derivatives of the marginal variance $\bm V$.  We can summarize 
the Fisher information matrix result for the LMM as: 
$$
        \bm A = \left[\begin{array}{ccc|ccc}
                   &&\\
                   & \bm X^{\top} \bm V^{-1} \bm X &&& \bm 0 &\\
                   &&\\
                   \hline
                   &&\\
                   & \bm 0 &&& \bm \left (\frac{1}{2}\right) \text{tr}
      \left [ \bm V^{-1} \left (\frac{\partial \bm V}
      {\partial \sigma_{k1}^2}\right)
      \bm V^{-1} \left (\frac{\partial \bm V}
      {\partial \sigma_{k2}^2}\right)\right ]&\\
                   &&\\
                   \end{array}\right].
$$
These results are
equivalent to Equations 6.69 to 6.74 of \cite{mcc01}.

We can then invert the information matrix to obtain the
variance-covariance matrix.  In the \code{vcov.lmerMod()} function 
from \pkg{merDeriv}, we use the default 
argument \code{full = TRUE} to get the variance-covariance matrix w.r.t.\ 
all parameters in the model.  If \code{full = FALSE}, the
variance-covariance matrix w.r.t.\ only fixed parameters is returned.  
To switch between the observed
and expected information matrix, we can supply the argument
\code{information = "observed"} or \code{information = "expected"}.  
The default option is ``expected'' due to its wider usage.  

\section[Computational relation]{Relation to \code{lmerMod} objects}
In this section, we describe how the quantities needed to
compute the scores, %(Equation~\ref{eq:scoresigma} and 
% Equation~\ref{eq:scoresbeta})
Hessian, and Fisher information matrix can be obtained from 
an \code{lmerMod} object.
The data and model matrices $\bm y$, $\bm X$, $\bm \beta$, and $\bm Z$ can
be obtained directly from \pkg{lme4} via \code{getME()}.
The only remaining components, then, are $\bm V$ and
$\partial \bm V/\partial \bm \sigma^2$.  In the following,
we focus on how to indirectly obtain these components.

In the \pkg{lme4} framework, the random effects covariance matrix
$\bm G$ is decomposed via \citep{lme4}:
\begin{equation}
  \label{eq:glme4}
  \bm G = \bm \Lambda_{\bm \theta} \bm \Lambda_{\bm \theta}^{\top} \sigma_r^2,
\end{equation}
where $\bm \Lambda_{\bm \theta}$ is a $q \times q$ lower diagonal matrix,
called the \emph{relative covariance factor}.  It can be seen as a
Cholesky decomposition of $\bm G/\sigma_r^2$.  The dimension of
$\bm \Lambda_{\bm \theta}$ is the same as that of $\bm G$.  Additionally,
the position of $\sigma_k^2$ in $\bm G$ is the same as the position
of $\theta_k$ in $\bm \Lambda_{\bm \theta}$.

Inserting Equation~\ref{eq:glme4} into Equation~\ref{eq:marginv},
we can express $\bm V$ as
\begin{equation}
  \label{eq:computev}
  \bm V = (\bm Z \bm \Lambda_{\bm \theta} \bm \Lambda_{\bm \theta}^{\top}
  \bm Z^{\top} + \bm I_n) \sigma_r^2.
\end{equation}
Equation~\ref{eq:computev} is mathematically equivalent to
Equation~\ref{eq:marginv}, but it has computational advantages when, e.g., 
a random effect variance is close to 0.  

Using Equation~\ref{eq:marginv}, the term
$\partial \bm V/\partial \sigma_k^2$ can usually be expressed as
\begin{equation}
  \label{eq:compusigma}
  \bm Z \frac{\partial \bm G}{\partial \sigma_k^2} \bm Z^{\top},
\end{equation}
so long as $\sigma_k^2$ is not the residual variance.
The partial derivative $\frac{\partial \bm G}{\partial \sigma_k^2}$ is 
then a matrix of the same dimension as $\bm G$, with an entry of 
$1$ corresponding to the location of $\sigma_k^2$ and $0$ elsewhere.

Because the location of $\sigma_k^2$ within $\bm G$ matches 
its location within $\bm \Lambda_{\bm \theta}$, we can use
$\bm \Lambda_{\bm \theta}$ to facilitate computation of
$\partial \bm V/\partial \sigma_k^2$.  The only trick is that 
$\bm G$ is symmetric, whereas $\bm \Lambda_{\bm \theta}$ is lower diagonal.

The code below illustrates implementation of this strategy, where 
\code{object} is a fitted model of class \code{lmerMod}.
We use \code{forceSymmetric()} to convert the lower
diagonal information from $\bm \Lambda_{\bm \theta}$ into the symmetric
$\bm G$.

\begin{Schunk}
\begin{Sinput}
R>   ## "object" is a fitted model of class lmerMod.
R>   parts <- getME(object, "ALL")
R>   uluti <- length(parts$theta)
R>   devLambda <- vector("list", uluti)
R>   devV <- vector ("list", (uluti + 1))
R> 
R>   ## get the position of parameters in Lambda matrix
R>   LambdaInd <- parts$Lambda
R>   LambdaInd@x[] <- as.double(parts$Lind)
R> 
R>   for (i in 1:uluti) {
+      devLambda[[i]] <- forceSymmetric(LambdaInd==i, uplo = "L")
+      devV[[i]] <- tcrossprod(tcrossprod(parts$Z, t(devLambda[[i]])), parts$Z)
+    }
\end{Sinput}
\end{Schunk}

Finally, for the derivative with respect to the residual variance, it
is obvious that
$\partial \bm V/\partial \sigma_r^2=\bm I_n$ so long
as $\bm R = \sigma_r^2 \bm I$ \cite[also see][p.\ 137]{stroup12}.
% This further assumes that
% $\bm G$ and $\bm R$ should be
% independent, regardless of the compuational expression of $\bm V$ in
% \pkg{lme4}.

The above results are sufficient for obtaining the derivatives necessary for 
computing the Huber-White sandwich estimator and for carrying out additional 
statistical tests (see the Discussion section). In the following sections, 
we will 
describe the Huber-White sandwich estimator for
linear mixed models with independent clusters, then provide an application. 

\section{Huber-White sandwich estimator}
Let $\bm y_{c_j}$ contain the observations within cluster $c_j$. If  
observations in different clusters are independent (as is the case in many 
linear mixed models), then we can write
\begin{equation}
  \label{eq:likelihood}
    \ell(\bm \sigma^2, \bm \beta; \bm{y}) = \sum_{j=1}^J \ell(\bm \sigma^2,
    \bm \beta; \bm y_{c_j}),
\end{equation}
where $J$ is the total number of clusters and $\ell()$ is defined in 
Equation~\ref{eq:obj}. The first and second partial derivatives of $\ell$ w.r.t.
$\bm \xi = (\bm \sigma^2\ \bm \beta)^\top$ can then be written as
\begin{eqnarray}
  \label{eq:firstder}
    \ell^{'}(\bm \xi; \bm y) &=& \sum_{j=1}^{J}
    \frac{\partial \ell(\bm \xi; \bm y_{c_j})}
    {\partial \bm \xi} = \sum_{j=1}^{J}
    \sum_{i \in c_j} s_i(\bm \xi; y_i)\\
  \label{eq:secondder}
    \ell^{''}(\bm \xi; \bm y) &=& \sum_{j=1}^{J}\frac{\partial^2
    \ell(\bm \xi; \bm y_{c_j})}
    {\partial \bm \xi^2},
\end{eqnarray}
where $\frac{\partial \ell(\bm \xi; \bm y_{c_j})}{\partial \bm \xi}$
represents the first derivative within cluster $c_j$, which can be 
expressed as the sum of the casewise score $s_i()$ belonging to 
$c_j$. The function $s_i()$ has also been studied in other 
contexts \citep[e.g.,][]{WanMerZei14, zeihor07}.

Inference about $\bm \xi$ relies on a central limit theorem:
\begin{equation}
  \label{eq:clt}
    \sqrt{J}(\hat{\bm \xi} - \bm \xi)\xrightarrow{d}
    N(\bm 0, \bm V(\bm \xi)),
\end{equation}
where $\xrightarrow{d}$ denotes convergence in distribution.   
The traditional estimate of $\bm V(\bm \xi)$ relies on
Equation~\ref{eq:secondder}, whereas
the Huber-White sandwich estimator of $\bm V(\bm \xi)$ is defined as
\citep[e.g.,][]{freed12, white80, sand2}:
\begin{equation}
  \label{eq:aba}
    \bm V(\hat{\bm \xi}) = (\bm A)^{-1}\bm B(\bm A)^{-1},
\end{equation}
where $\bm A=-E(\ell^{''}(\hat{\bm \xi}); \bm{y})$ and
$\bm B=\text{Cov}(\ell^{'}(\hat{\bm \xi}; \bm{y}))$.  
The square roots of the diagonal
elements of $\bm V$ are the ``robust standard errors.''
% In most models, $\bm V(\bm \theta)$ can be simplified
% as $(-\bm A)$, which is the Hessian available in most packages.

When the model is correctly specified, the Huber-White sandwich estimator
corresponds to the Fisher information matrix.  However, the estimator 
is often used in non-\emph{i.i.d.} samples to ``correct'' the 
information matrix for misspecification \citep[e.g.,][]{freed12}.  
While mixed models explicitly handle lack of independence via 
random effects, the Huber-White estimators can still be applied to 
these models to address remaining model misspecifications such as outliers in 
random effects or deviations from normality \citep{roblmm, kol13}.

To construct the Huber-White sandwich estimator,
$\bm A$ can be obtained from Equation~\ref{eq:secondder}, whose analytic 
expression for the linear mixed model is expressed in Section 3.2. The matrix
$\bm B$ can then be constructed via \citep[e.g.,][]{freed12}:
\begin{equation}
  \label{eq:clustermeat}
     \bm B=\sum_{j=1}^{J}\left[\sum_{i \in c_j} s_i(\bm \xi; y_i) \right ]^{\top}
      \left [\sum_{i \in c_j} s_i(\bm \xi; y_i)\right ].
\end{equation}
Thus, we require the derivations presented in the previous section: the 
``score'' terms $s_i(\bm{\xi}; y_i)$ $(i=1,\ldots,n)$ and the
information matrix using the marginal
likelihood from Equation~\ref{eq:obj}.

\section{Application}
In this section, we illustrate how the package can be used to obtain
clusterwise robust standard errors for the \code{sleepstudy}
data \citep{belenky03} included in \pkg{lme4}.  This dataset includes
$18$ subjects participating in a sleep deprivation study, where each
subject's reaction time was monitored for $10$ consecutive days.
The reaction times are nested by subject and continuous in measurement,
hence the linear mixed model.

We first load package \pkg{lme4}, along with the \pkg{merDeriv} package 
that is the focus of this paper.

\begin{Schunk}
\begin{Sinput}
R> library("lme4")
R> library("merDeriv")
\end{Sinput}
\end{Schunk}

Next, we fit a model with \code{Days} as the covariate, including random
intercept and slope effects that are allowed to covary.  There are six
free model parameters: the fixed intercept and slope $\beta_0$ and $\beta_1$,
the random variance and covariances $\sigma_0^2$, $\sigma_1^2$,
and $\sigma_{01}$, and the residual variance $\sigma_r^2$.

\begin{Schunk}
\begin{Sinput}
R> lme4fit <- lmer(Reaction ~ Days + (Days|Subject), sleepstudy, 
+    REML = FALSE)
\end{Sinput}
\end{Schunk}

This particular model can also be estimated as a structural equation
model via package \pkg{lavaan}, facilitating the comparison of our results 
with a benchmark.  We first convert the data to wide format and then 
specify/estimate the model:

\begin{Schunk}
\begin{Sinput}
R> testwide <- reshape2::dcast(sleepstudy, Subject ~ Days, 
+    value.var = "Reaction")
R> names(testwide)[2:11] <- paste("d", 1:10, sep = "")
R> ## describe latent model
R> latent <- 'i =~ 1*d1 + 1*d2 + 1*d3 + 1*d4 + 1*d5
+    + 1*d6 + 1*d7 + 1*d8 + 1*d9 + 1*d10
+  
+    s = ~ 0*d1 + 1*d2 + 2*d3 + 3*d4 + 4*d5
+    + 5*d6 + 6*d7 + 7*d8 + 8*d9 + 9*d10
+  
+    d1 ~~ evar*d1
+    d2 ~~ evar*d2
+    d3 ~~ evar*d3
+    d4 ~~ evar*d4
+    d5 ~~ evar*d5
+    d6 ~~ evar*d6
+    d7 ~~ evar*d7
+    d8 ~~ evar*d8
+    d9 ~~ evar*d9
+    d10 ~~ evar*d10
+  
+    ## reparameterize as sd
+    sdevar := sqrt(evar)
+    i ~~ ivar*i
+    isd := sqrt(ivar)'
R> ## fit model in lavaan
R> lavaanfit <- growth(latent, data = testwide, estimator = "ML")
\end{Sinput}
\end{Schunk}

The parameter estimates from the two packages (not shown) all agree to at 
least three decimal places. Below, we examine the agreement of derivative computations.

\subsubsection{Scores}
The analytic casewise and clusterwise scores are obtained
via \code{estfun.lmerMod()}, using the arguments \code{level = 1} and
\code{level = 2}, respectively.  The sum of scores (either casewise or 
clusterwise) equals the gradient, which is close to zero at the ML estimates.
\begin{Schunk}
\begin{Sinput}
R> score1 <- estfun.lmerMod(lme4fit, level = 1)
R> gradients1 <- colSums(score1)
R> gradients1
\end{Sinput}
\begin{Soutput}
                 (Intercept)                         Days 
                    2.39e-14                     2.38e-13 
     cov_Subject.(Intercept) cov_Subject.Days.(Intercept) 
                    2.94e-09                     4.19e-08 
            cov_Subject.Days                     residual 
                    8.29e-08                    -7.38e-09 
\end{Soutput}
\end{Schunk}

\begin{Schunk}
\begin{Sinput}
R> score2 <- estfun.lmerMod(lme4fit, level = 2)
R> gradients2 <- colSums(score2)
R> gradients2
\end{Sinput}
\begin{Soutput}
                 (Intercept)                         Days 
                    2.39e-14                     2.38e-13 
     cov_Subject.(Intercept) cov_Subject.Days.(Intercept) 
                    2.94e-09                     4.19e-08 
            cov_Subject.Days                     residual 
                    8.29e-08                    -7.38e-09 
\end{Soutput}
\end{Schunk}

\begin{figure}
\caption{Comparison of scores obtained via \code{estfun.lavaan}
  and \code{estfun.lmerMod}. In the left panel, 
  the y-axis represents analytic, clusterwise scores
  obtained from \code{estfun.lmerMod}, and the x-axis represents
  clusterwise scores obtained from \code{estfun.lavaan}. The dashed
  line serves as a reference line as y = x; In the right panel, 
  the y-axis represents the difference between the scores obtained via
  \code{estfun.lavaan} and \code{estfun.lmerMod}, and the x-axis represents
  the clusterwise scores obtained from \code{estfun.lavaan}. The dashed
  line serves as a reference line as y = 0;}
\label{fig:scorelavaan}
\begin{Schunk}

{\centering \includegraphics[width=6.5in,height=4.5in]{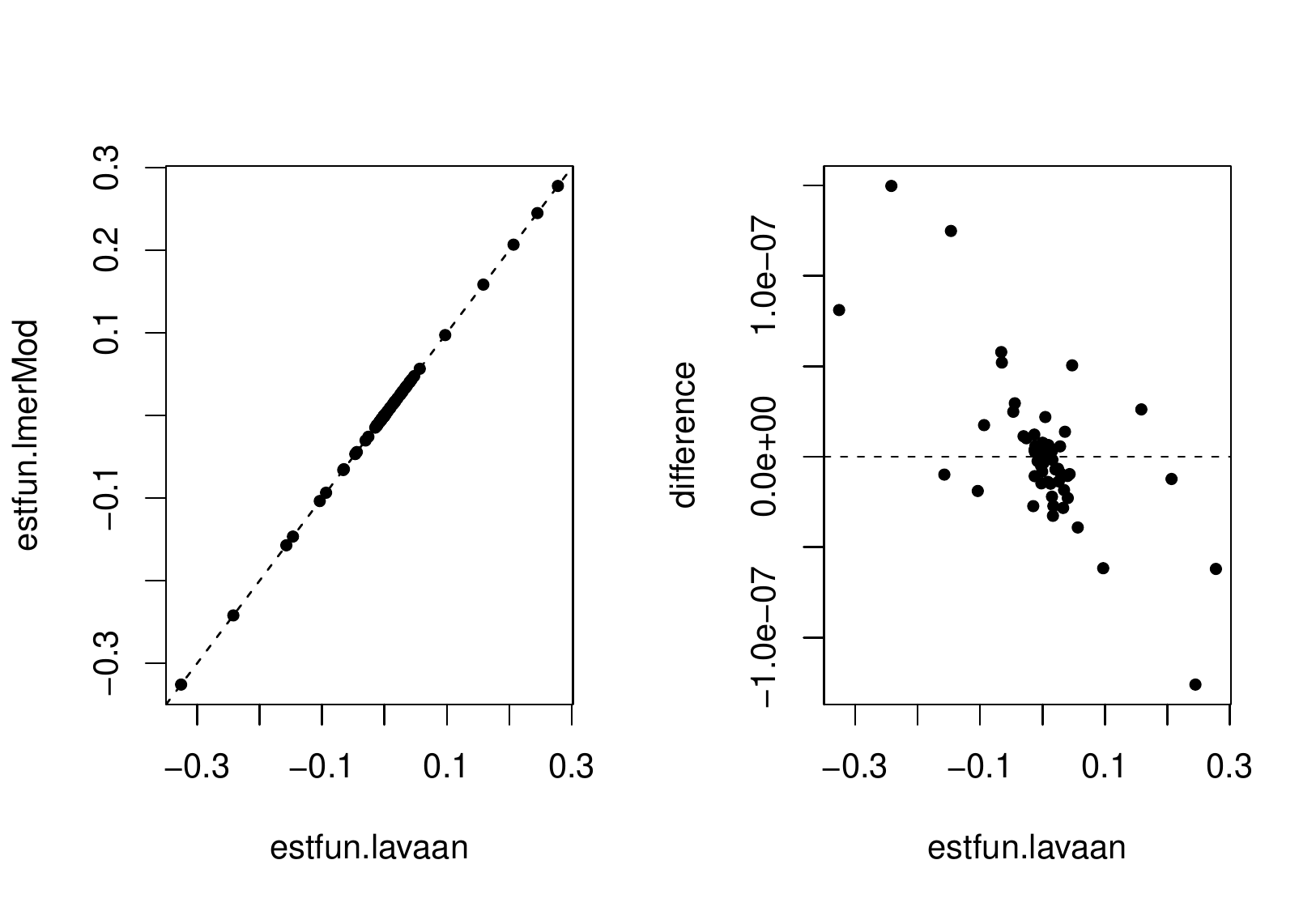} 

}

\end{Schunk}
\end{figure}

The clusterwise scores are also provided by \code{estfun.lavaan()} in
\pkg{lavaan}. Figure~\ref{fig:scorelavaan} presents a comparison
between the clusterwise scores obtained
from \code{estfun.lmerMod()} and \code{estfun.lavaan()},
showing they are nearly identical.  The absolute difference between the scores
obtained from these two packages is within $1.5 \times 10^{-7}$.  The 
sum of squared differences is within $2.2 \times 10^{-14}$.

%The resulting \code{meat} (clusterwise) is as following:

\subsubsection{Variance covariance matrices}
We also compare the variance covariance matrix calculated via our \pkg{lme4}
second derivatives to the \code{vcov()} output of lavaan.
The results are displayed in Table~\ref{tab:bread}.  The maximum of the
absolute difference for all components in the variance covariance matrix is
0.07.  This minor
difference is due to the fact that 
\pkg{lavaan} applies the delta method to compute the Fisher 
information matrix for
defined parameters \citep{rosv12,obe14}.  In contrast, 
\pkg{merDeriv} utilizes analytic expressions.  This 
difference is ignorable due to the small relative 
difference (within $10^{-6}$).

% latex table generated in R 3.4.0 by xtable 1.8-2 package
% Thu Nov  2 13:50:28 2017
\begin{table}[ht]
\centering
\scalebox{0.75}{
\begin{tabular}{llllll}
  \hline
Column name & Row name & merDeriv & lavaan & Abs Diff & Relative Diff \\ 
  \hline
(Intercept) & (Intercept) &    43.99 &    43.99 & 0.00 & 0.00 \\ 
  Days & (Intercept) &    -1.37 &    -1.37 & 0.00 & 0.00 \\ 
  cov\_Subject.(Intercept) & (Intercept) &     0.00 &     0.00 & 0.00 & -- \\ 
  cov\_Subject.Days.(Intercept) & (Intercept) &     0.00 &     0.00 & 0.00 & -- \\ 
  cov\_Subject.Days & (Intercept) &     0.00 &     0.00 & 0.00 & -- \\ 
  residual & (Intercept) &     0.00 &     0.00 & 0.00 & -- \\ 
  (Intercept) & Days &    -1.37 &    -1.37 & 0.00 & 0.00 \\ 
  Days & Days &     2.26 &     2.26 & 0.00 & 0.00 \\ 
  cov\_Subject.(Intercept) & Days &     0.00 &     0.00 & 0.00 & -- \\ 
  cov\_Subject.Days.(Intercept) & Days &     0.00 &     0.00 & 0.00 & -- \\ 
  cov\_Subject.Days & Days &     0.00 &     0.00 & 0.00 & -- \\ 
  residual & Days &     0.00 &     0.00 & 0.00 & -- \\ 
  (Intercept) & cov\_Subject.(Intercept) &     0.00 &     0.00 & 0.00 & -- \\ 
  Days & cov\_Subject.(Intercept) &     0.00 &     0.00 & 0.00 & -- \\ 
  cov\_Subject.(Intercept) & cov\_Subject.(Intercept) & 70366.08 & 70366.15 & 0.07 & 0.00 \\ 
  cov\_Subject.Days.(Intercept) & cov\_Subject.(Intercept) & -2282.47 & -2282.46 & 0.01 & 0.00 \\ 
  cov\_Subject.Days & cov\_Subject.(Intercept) &    92.56 &    92.56 & 0.00 & 0.00 \\ 
  residual & cov\_Subject.(Intercept) & -2058.08 & -2058.08 & 0.00 & 0.00 \\ 
  (Intercept) & cov\_Subject.Days.(Intercept) &     0.00 &     0.00 & 0.00 & -- \\ 
  Days & cov\_Subject.Days.(Intercept) &     0.00 &     0.00 & 0.00 & -- \\ 
  cov\_Subject.(Intercept) & cov\_Subject.Days.(Intercept) & -2282.47 & -2282.46 & 0.01 & 0.00 \\ 
  cov\_Subject.Days.(Intercept) & cov\_Subject.Days.(Intercept) &  1838.33 &  1838.33 & 0.00 & 0.00 \\ 
  cov\_Subject.Days & cov\_Subject.Days.(Intercept) &  -115.28 &  -115.28 & 0.00 & 0.00 \\ 
  residual & cov\_Subject.Days.(Intercept) &   324.96 &   324.96 & 0.00 & 0.00 \\ 
  (Intercept) & cov\_Subject.Days &     0.00 &     0.00 & 0.00 & -- \\ 
  Days & cov\_Subject.Days &     0.00 &     0.00 & 0.00 & -- \\ 
  cov\_Subject.(Intercept) & cov\_Subject.Days &    92.56 &    92.56 & 0.00 & 0.00 \\ 
  cov\_Subject.Days.(Intercept) & cov\_Subject.Days &  -115.28 &  -115.28 & 0.00 & 0.00 \\ 
  cov\_Subject.Days & cov\_Subject.Days &   184.21 &   184.21 & 0.00 & 0.00 \\ 
  residual & cov\_Subject.Days &   -72.21 &   -72.21 & 0.00 & 0.00 \\ 
  (Intercept) & residual &     0.00 &     0.00 & 0.00 & -- \\ 
  Days & residual &     0.00 &     0.00 & 0.00 & -- \\ 
  cov\_Subject.(Intercept) & residual & -2058.08 & -2058.08 & 0.00 & 0.00 \\ 
  cov\_Subject.Days.(Intercept) & residual &   324.96 &   324.96 & 0.00 & 0.00 \\ 
  cov\_Subject.Days & residual &   -72.21 &   -72.21 & 0.00 & 0.00 \\ 
  residual & residual &  5957.61 &  5957.61 & 0.00 & 0.00 \\ 
   \hline
\end{tabular}
}
\caption{Comparison between \pkg{merDeriv} 
  \code{vcov.lmerMod()} output and \pkg{lavaan} \code{vcov()} 
  output for the \code{sleepstudy} data.  The first two columns 
  describe the specific matrix entry being compared, the third 
  and fourth columns show the estimates, the fifth and sixth 
  column shows the absolute and relative difference.} 
\label{tab:bread}
\end{table}

Finally, the clusterwise Huber-White sandwich estimator is shown in
Table~\ref{tab:sandwich},
which is comparable to the one provided by \pkg{lavaan}.
The maximum of the absolute difference for all
components in the variance covariance matrix is
0.05.  The
minor difference is again caused by the aforementioned reasons.

% latex table generated in R 3.4.0 by xtable 1.8-2 package
% Thu Nov  2 13:50:28 2017
\begin{table}[ht]
\centering
\scalebox{0.75}{
\begin{tabular}{llllll}
  \hline
Column name & Row name & merDeriv & lavaan & Abs Diff & Relative Diff \\ 
  \hline
(Intercept) & (Intercept) &     43.99 &     43.99 & 0.00 & 0.00 \\ 
  Days & (Intercept) &     -1.37 &     -1.37 & 0.00 & 0.00 \\ 
  cov\_Subject.(Intercept) & (Intercept) &   -523.40 &   -523.41 & 0.01 & 0.00 \\ 
  cov\_Subject.Days.(Intercept) & (Intercept) &    -20.77 &    -20.77 & 0.00 & 0.00 \\ 
  cov\_Subject.Days & (Intercept) &     -5.92 &     -5.92 & 0.00 & 0.00 \\ 
  residual & (Intercept) &    149.15 &    149.15 & 0.00 & 0.00 \\ 
  (Intercept) & Days &     -1.37 &     -1.37 & 0.00 & 0.00 \\ 
  Days & Days &      2.26 &      2.26 & 0.00 & 0.00 \\ 
  cov\_Subject.(Intercept) & Days &    -56.09 &    -56.09 & 0.00 & 0.00 \\ 
  cov\_Subject.Days.(Intercept) & Days &      0.18 &      0.18 & 0.00 & 0.00 \\ 
  cov\_Subject.Days & Days &     -1.98 &     -1.98 & 0.00 & 0.00 \\ 
  residual & Days &     78.71 &     78.71 & 0.00 & 0.00 \\ 
  (Intercept) & cov\_Subject.(Intercept) &   -523.40 &   -523.41 & 0.01 & 0.00 \\ 
  Days & cov\_Subject.(Intercept) &    -56.09 &    -56.09 & 0.00 & 0.00 \\ 
  cov\_Subject.(Intercept) & cov\_Subject.(Intercept) &  45232.13 &  45232.18 & 0.05 & 0.00 \\ 
  cov\_Subject.Days.(Intercept) & cov\_Subject.(Intercept) &   1055.38 &   1055.38 & 0.00 & 0.00 \\ 
  cov\_Subject.Days & cov\_Subject.(Intercept) &    427.39 &    427.39 & 0.00 & 0.00 \\ 
  residual & cov\_Subject.(Intercept) & -27398.62 & -27398.62 & 0.00 & 0.00 \\ 
  (Intercept) & cov\_Subject.Days.(Intercept) &    -20.77 &    -20.77 & 0.00 & 0.00 \\ 
  Days & cov\_Subject.Days.(Intercept) &      0.18 &      0.18 & 0.00 & 0.00 \\ 
  cov\_Subject.(Intercept) & cov\_Subject.Days.(Intercept) &   1055.38 &   1055.38 & 0.00 & 0.00 \\ 
  cov\_Subject.Days.(Intercept) & cov\_Subject.Days.(Intercept) &   1862.99 &   1862.99 & 0.00 & 0.00 \\ 
  cov\_Subject.Days & cov\_Subject.Days.(Intercept) &    -89.28 &    -89.28 & 0.00 & 0.00 \\ 
  residual & cov\_Subject.Days.(Intercept) &   1214.37 &   1214.37 & 0.00 & 0.00 \\ 
  (Intercept) & cov\_Subject.Days &     -5.92 &     -5.92 & 0.00 & 0.00 \\ 
  Days & cov\_Subject.Days &     -1.98 &     -1.98 & 0.00 & 0.00 \\ 
  cov\_Subject.(Intercept) & cov\_Subject.Days &    427.39 &    427.39 & 0.00 & 0.00 \\ 
  cov\_Subject.Days.(Intercept) & cov\_Subject.Days &    -89.28 &    -89.28 & 0.00 & 0.00 \\ 
  cov\_Subject.Days & cov\_Subject.Days &    137.89 &    137.89 & 0.00 & 0.00 \\ 
  residual & cov\_Subject.Days &   -492.56 &   -492.56 & 0.00 & 0.00 \\ 
  (Intercept) & residual &    149.15 &    149.15 & 0.00 & 0.00 \\ 
  Days & residual &     78.71 &     78.71 & 0.00 & 0.00 \\ 
  cov\_Subject.(Intercept) & residual & -27398.62 & -27398.62 & 0.00 & 0.00 \\ 
  cov\_Subject.Days.(Intercept) & residual &   1214.37 &   1214.37 & 0.00 & 0.00 \\ 
  cov\_Subject.Days & residual &   -492.56 &   -492.56 & 0.00 & 0.00 \\ 
  residual & residual &  43229.03 &  43229.03 & 0.00 & 0.00 \\ 
   \hline
\end{tabular}
}
\caption{Comparison of the \code{sleepstudy} sandwich 
  estimator obtained from our \pkg{merDeriv} code with the 
  analogous estimator obtained from \pkg{lavaan}. The first two 
  columns describe the specific matrix entry being compared, the 
  third and fourth columns show the estimates, the fifth and sixth 
  column shows the absolute and relative difference.} 
\label{tab:sandwich}
\end{table}

% \subsection{Cross random effect model}
% \pkg{lme4} is particularly effective in fitting models with cross random effect
% where data were cross-classifed.  For example, in \code{Penicillin}
% data included in \pkg{lme4} package, penicillin's
% diameter were measured in different samples
% and plates \citep{davies1972}.  In other word, any single penicillin can be
% classied according to its source sample or the plate it reside in.
% In this particular
% dataset, there are $24$ plates, $6$ samples and a total of $144$
% penicillin's diameters being measured. We fit
% a cross random effect model as described in \cite[Chapter 2]{bates2010book}.

% Since the cross random effect was handled in $\bm Z$ and
% $\bm \Lambda_{\bm \theta}$ as documented in \cite{bates2010book},
% using the current
% \code{bread}, \code{meat} and \code{sandwich}, we can automatically get
% the corresponding matrix based on sample and plate as shown below.

% <<crossed random effect, echo=TRUE>>=
% fm2 <- lmer(diameter ~ 1 + (1|plate) + (1|sample), Penicillin, REML=FALSE)
% sandwich(fm2, level="level2")
% @

% \fixme{the std provided is not based on clusters.}

\section{Discussion}
In this paper, we illustrated how to obtain the Huber-White sandwich
estimator of estimated parameters arising from objects of
class \code{lmerMod} with independent clusters.  This required us to derive
observational (and clusterwise) scores for fixed and random
parameters (leading to the ``meat'') as well as a Fisher information matrix
that included random effect variances and
covariances (leading to the ``bread'').  In the discussion below, 
we address extensions to related 
statistical metrics and models.

\subsection{Restricted maximum likelihood (REML)}

While we focused on linear mixed models estimated via maximum 
likelihood (ML), extension to restricted maximum likelihood (REML) is 
straightforward.  
The central idea of REML is to maximize the likelihood function of 
variance parameters after 
accounting for the fixed effects.  By using this approach, the downward 
bias of 
ML for variance estimates can be eliminated (similarly to division 
of $n$ versus $n-1$ in simple variance calculations), so REML is used often in 
LMM applications \citep{stroup12}. 

Referring to the \code{sleepstudy} example, package \pkg{merDeriv} can 
provide scores (\code{estfun}) 
and variance covariance matrix (\code{vcov}) based on the REML likelihood 
function and corresponding estimates.  
The fixed effects parameters are equivalent for ML and REML, whereas the 
corresponding \code{vcov} components are larger based on REML.  For example, 
the ML variance for the estimated fixed intercept is 
43.99 
whereas the REML variance for the estimated fixed intercept is  
46.57. 

\subsection{Statistical tests}
The scores derived in this paper can potentially be used to carry out
a variety of score-based statistical tests.  For example, the
``fluctuation test'' framework discussed by \cite{zeihor07}, \cite{MerZei13}, 
and others generalizes the traditional
score (Lagrange multiplier) test and is used to detect parameter instability 
across orderings of observations. The tests have been critical for the
development of model-based recursive partitioning procedures available
via packages such as \pkg{partykit} \citep{party}.

The code that we present here facilitates application of score-based
tests to linear mixed models, because the tests described in the
previous paragraph are available via object-oriented \proglang{R} packages.
That is to say the aforementioned packages can be applied to  
linear mixed models estimated via \pkg{lme4}, because we have supplied the generic function 
\code{estfun} for models of class \code{lmerMod}. 
A challenge involves the fact that much of the above theory
requires observations to be independent.  For the linear mixed models with 
independent clusters, tests can often be applied immediately. %  However, 
% compared with the statistical tests based on traditional linear model with 
% $n$ \emph{i.i.d.} observations, the tests based on linear mixed model of 
% the same observation
% will generally have lower power.  The reason is that the central limit theory 
% shown in Equation~\ref{eq:clt} involves the number of clusters $J$ instead of 
% the number of total observation $n$. 
However, while we can
test parameter instability across independent clusters, it is
more difficult to test for instability across correlated observations within a
cluster. A related issue, further described below, arises when we attempt to 
apply sandwich estimators to models with crossed random effects.

\subsection{Models with multiple random effects terms}
The ``independence'' challenges described in the previous section
translate to the setting of models with multiple random effects terms, such as 
(partially) crossed
random effects or models with multilevel nested 
designs \cite[e.g., three-level models;][Ch.2]{bates10}. 
These correspond to situations where there are at
least two unique variables defining clusters (for example, clusters
defined by primary school attended and by secondary school attended).
In this case, we cannot simply sum scores within a cluster
to obtain independent, clusterwise scores. This is because observations in
different clusters on the first grouping variable may be in
the same cluster on the second grouping variable. Thus, it is unclear
how the statistical machinery developed for independent
observations (e.g., robust standard errors, instability tests) can transfer to
models with partially crossed random effects. While 
our \code{estfun.lmerMod()} code can return casewise
scores and \code{vcov.lmerMod()} can
return the full variance covariance matrix of all \code{lmerMod} objects, 
it is unclear how to further use these results. 

The main difficulty involves construction of
the ``meat'', which is the variance of the first derivatives based on the 
grouping variable.  One possible solution is to create separate ``meats'' 
based on 
different grouping variables, accounting for covariances between the meats. %  The experimental design information, such as cross random, 
% partially cross random
% or nested random, is contained in the grouping vector 
% for each observation.
This approach is described in \cite{ras94} 
and \cite{cam11} to decompose 
parameter variances when there are multiple grouping variables.  
It may be possible to apply the same idea to our problem, and we plan to study this in the future. 

\subsection{GLMM}
Finally, the procedures described here for scores, Hessians, Fisher information
and sandwich estimators can be extended to generalized linear mixed
models estimated via \code{glmer()}.
The technical difficulty involved with this extension is the observational
scores. In the linear mixed model, we can derive the analytic scores for each
observation because we know that the marginal distribution is normal.
In the GLMM, the marginal distribution is typically unknown, and we require
integral approximation methods (e.g., quadrature or the Laplace approximation) 
to obtain the scores and second derivatives.
Combination of these integral approximation methods with the
\pkg{lme4} penalized least squares approach presents a challenge
that we have not yet overcome. We plan to do so in the future.

\section{Acknowledgments}
This research was partially supported by NSF grant 1460719. 

\bibliography{refs}
\end{document}